\documentstyle[12pt,epsf]{article}

\newcommand{\eq}[1]{equation~(\ref{#1})}
\newcommand{\beq}{\begin{equation}}
\newcommand{\eeq}{\end{equation}}

\begin{document}
\def\lag{\langle}
\def\rag{\rangle}

\title{Multimagnetical Simulations$^1$}
\footnotetext[1]{\em To appear in the 
Proceedings of the Workshop on ``Recent Developments in
 Computer Simulation Studies in Condensed Matter Physics," 
 Athens Ga, February 16--21, 1992.}

\author{U.\ Hansmann$^{2,3}$,\ B.A.\ Berg$^{2,3}$ and T.\  Neuhaus$^4$}
\footnotetext[2]{{\em
 Department of Physics, The Florida State University, Tallahassee,
 FL 32306, USA}}
\footnotetext[3]{{\em
 Supercomputer Computations Research Institute Tallahassee, The Florida
 State University, Tallahassee, FL 32306, USA}}
\footnotetext[4]{{\em
 Fakult\"at f\"ur Physik, Universit\"at Bielefeld,
 D-4800 Bielefeld, FRG}}

\maketitle
\pagestyle{empty}
\begin{abstract}
 We modified the recently proposed multicanonical MC algorithm
for the case of a magnetic field driven order--order phase transition. 
We test this {\em multimagnetic} Monte Carlo algorithm 
for the $D=2$ Ising model at $\beta=0.5$ and simulate square lattices 
up to size $100 \times 100$. On these lattices with periodic boundary 
conditions  it is possible to enhance the appearance of order-order 
interfaces during the simulation by many orders of magnitude as compared 
to the standard Monte Carlo simulation.
\end{abstract}

   Interfaces between ordered and disordered and, correspondingly, between
ordered and ordered physical states play an important role in statistical
mechanics and field theoretic models of fundamental gauge
interactions. 
   Past numerical studies of the properties of interfaces were, however,
hampered by a problem of principle. The surface tension per unit area $F^s$
between different states 
has a finite value. Thus in the canonical ensemble, where configurations
are sampled with the Boltzmann weights $\cal{P}^B \propto$ $e^{-\beta H}$,
configurations containing interfaces with an area $A$  
are suppressed by exponentially small factors $e^{- A F^s}$. Corresponding
to this is the exponentially fast increase of the
tunneling time with the area $A$ between pure phases of the system, when
the system is simulated with {\em local} Monte Carlo (MC) algorithms.
To overcome this problem  for a temperature driven, order--disorder
first order phase transition 
two of the present authors recently proposed a {\it
multicanonical} Monte Carlo algorithm \cite{our1,our2}. In \cite{BHN} this 
algorithm is generalized for the case of a magnetic field driven
order--order phase transition. Here we present this {\it multimagnetic} Monte
Carlo algorithm and illustrate how things work by simulating the $2D$ Ising
model deep in the broken region.

For our model, spins $s_i = \pm 1$ are defined on  sites of
a square lattice of volume $V = L\times L$ with periodic boundary
conditions and the symbol $<i,j>$ is used to denote nearest neighbours. The
partition function of the $2D$ Ising model is given by 
\begin{equation}  
Z\ = \ \sum_{\rm configurations} \exp ( - \beta H) , 
\end{equation}
where the Ising Hamiltonian $H$ \
\begin{equation}
H\ =\ H_I\ -\ h M ~~~{\rm with}~~~ H_I = - \sum_{<i,j>} s_i s_j
~~{\rm and}~~ M = \sum_i s_i
\end{equation}
contains the nearest neighbour interaction term $H_I$ and a term
which couples the magnetic field $h$ to the magnetization
$M=\sum_i s_i$. For this model the exact planar interface surface tension is
known to be given by \cite{Onsager}
\begin{equation}
F^s\ =\ 2\beta - \ln \left[ {1+e^{-2\beta} \over 1-e^{-2\beta}} \right]
~~{\rm for}~~ \beta\ge\beta_c = {1\over2} \ln (1+\sqrt{2}) = 0.4406...~.
\end{equation}

We chose $\beta = 0.5$. To demonstrate the problem let us first look at
the probability distributions of the magnetization $M$ (see figure 1).
\begin{figure}
\vspace{9cm}
\caption[fig1]{Boltzmann probability distributions $P_L (M)$ for the
magnetization on lattices with  $L=10$ to $L=100$. 
We have adopted the normalization $\sum_{M} P_{L} (M) = 1$ } 
\end{figure} 
\noindent The distributions are sharply double peaked and we denote the positions of
the maxima by $\pm M^{\max}_L$. As the model is
globally $Z(2)$ symmetric
we know
$P_L (M) = P_L (-M)$.
 The distribution
takes its minimum at $P_L^{\min}=P_L (0)$.
The logarithmic scale of figure~1 displays that
more than twenty orders of magnitude are involved, i.e.
$P_L^{\min} / P_L^{\max} < 10^{-20}$ for our largest lattice with $L=100$.
The standard MC algorithm would only sample configurations corresponding
to $P^{\min}_{100}$ if one could generate of the order $O(10^{20})$
or more statistically independent configurations.
 
 Here we overcome this difficulty by sampling
configurations with a multimagnetical weight factor
\begin{equation}
{\cal P}^{mm}_L (M)\ \sim\ \exp ( \alpha_L^k + h_L^k \beta M - \beta H_I )
~~~{\rm for}~~~ M_L^k < M \le M_L^{k+1}
\label{mumag}
\end{equation}
instead of sampling with the usual Boltzmann factor ${\cal P}^B$.
In \cite{BHN} we explain explicitly  how to calculate the $\alpha_L^k$ 
and $  h_L^k$ such that the multimagnetical probability distribution
${\cal P}^{mm}_L (M)$ becomes arbitrarily flat. 
The probability distribution $P_L (M)$
corresponding to the Boltzmann weight is then obtained
from ${\cal P}_L (M)$ by reweighting with
$\exp~(-\alpha^k_L~-~h^k_L~\beta_M)$~\cite{our1}.

Our simulations were performed on square lattices with linear size $L = 2$
up to $L = 100$. Up to $L \leq 16$  we get the parameters in
equation (\ref{mumag})
from standard heat bath simulations. This is no longer possible for
larger lattices and in this case, we chose them by making 
a finite size scaling (FSS)
prediction of the quantity $P_L(M)$ from the smaller systems everytime.
To
optimize the parameters we performed a second run and in some cases several 
more
runs to control our results. Our statistics for these investigations were
$4 \cdot 10^6$ sweeps per run and lattice size for $L = 2$ up to $L = 50$
and $8 \cdot 10^6$
sweeps for the larger systems. 200,000 additional, initial sweeps were performed
for reaching equilibrium with respect to the multimagnetical distribution.
One sweep updates every spin on the lattice once.

\begin{figure}
\vspace{9cm}
\caption[fig2]{\small
Tunneling times versus lattice size. The upper data are for
 the standard heat bath algorithm and the broken line extrapolates by means of 
a fit into the region where no data exist. The lower 
data points are obtained with our multimagnetical algorithm. }
\end{figure}
To compare the efficiency of our method with standard MC we measured the
tunneling time $\tau^t_L$ .  We define the
tunneling time $\tau_L^t$ as the average number of sweeps needed to get
from a configuration with magnetization $M=-M^{\max}_L$ to a
configuration with $M=M_L^{\max}$ {\bf and} back\cite{our2}.
In figure~2 we display on a log--log scale both the tunneling times for
the multimagnetic MC algorithm and the heatbath algorithm.  While 
there is an exponential fast increase of $\tau_L^t$ for the heatbath
algorithm, the increase of $\tau_L^t$ is for the multimagnetic algorithm
of the type of a power law divergence. 
 The ratio
\begin{equation}
 R\ =\ \tau_L^t(\rm heat bath) ~/~ \tau_L^t (\rm multicanonical) 
\end{equation}
is a direct measure for the improvement due to our method.
$R$ increases from a factor $4$ for
the smallest lattice ($L = 2$) up to $R \approx 450 $ for $L = 16$, the
biggest lattice size where it was with our statistics possible to get
data from standard MC. The extrapolation to $L = 100$ yields
$R \approx 6.1\times 10^{15}$, i.e. an improvement by more than fifteen 
orders of magnitude.
\begin{figure}
\vspace{9cm}
\caption[fig3]{Infinite volume estimate of the interfacial tension by means of
\eq{tens2}.}
\end{figure}

Former attempts 
\cite{Bind1,Bind2} to estimate the interface tension by a
finite size scaling ansatz failed for our $\beta$, deep in the broken
phase. In contradiction to this our multimagnetic data allows a highly
precise determination of the interface tension per unit area even in this
case. Following Binder \cite{Bind2}, the interface tension $F^s$ can be 
defined from the infinite volume limit of the quantity of 
$P_L^{\min}/P_L^{\max}$.
On finite lattices we define the effective interface tension by means of
\begin{equation}
F^s_L\ =\  {1\over 2 L} \ln \left( {P^{\max}_L \over P^{\min}_L} \right) .
\label{tens2}
\end{equation}

We estimate $F^s$ by mean of the FSS extrapolation towards $L=\infty$ (see
figure~3):
\begin{equation}
F^s_L\ =\ F^s + {a \over L} .
\end{equation}
 We  get a self-consistent fit for the range $L=30-100$ 
with the estimate 
$F^s = 0.2281~(8)$, which is in excellent agreement with the exact value
$F^s = 0.22806\ldots$.

It is worthwhile to inspect the distributions $P_L (M)$ (as plotted in
figure~1)  more closely.
As the lattice size increases we observe a flattening
of the logarithm of the probability distribution
close to values of the magnetization $m=0$
and it appears that in the thermodynamic limit the probability
distribution becomes constant in a finite range of mean
magnetization with $|m| \le m^c$. 
Such a behaviour is expected on the basis of a droplet
model~\cite{Bind1,Sloshman}.
In the region $m\le m^c$ the minimum free energy excess is dominated by a
single rectangular domain, extending throughout the lattice by making use
of its periodic boundary conditions. As the size of this block may be
varied without changing the free energy, one expects a
flat region.  To illustrate the droplet model,
we decided to write configurations on the computer disk which describe 
a tunneling transition on our $100 \times 100$ lattice . We
depict some examples in figures~4a-4d. On a qualitative level we find
support for the droplet model. Figure~4a depict a fairly ordered configuration
with magnetization $m=-0.9$. With increasing $m$ a single droplet
cluster 
begins to dominate the configuration, figure 4b for $m=-0.6$. This cluster 
grows  until it extends throughout the lattice by 
means of the periodic boundary conditions, figure~4c for $m=-0.2$.
Little changes are then observed in the flat region down to $m=0$(figure~4d). 
With further increase in $m$ ($m\ge 0$), similar pictures, 
with black and white interchanged, are seen to emerge. In order to gain a 
visual impression, we produced with this data a short video which shows 
this behaviour much more explicitly.

\begin{figure}
\epsffile{ising.eps}
\caption[fig4]{Configuration at $m=-0.9$ (a), $m=-0.6$ (b),
               $m=-0.2$ (c) and $m=0$ (d) on the 
               $100 \times 100$ lattice.}
\end{figure}

   In summary, we have shown that the multimagnetic  Monte Carlo algorithm
 eliminates the (with area $A$ of the interfaces) exponentially fast increase
of the tunneling time between the ordered phases of the system.
 The remaining
critical slowing down is of the type of a power law divergence.
Due to our new method it is possible to explore
configurations in phase space, which in the canonical ensemble are suppressed
by many orders of magnitude. 
We have performed a finite size scaling analysis
of the finite volume estimates of the interface tension and our
infinite volume value for the interface tension agrees 
with the analytically known Onsager result.

\section*{Acknowledgements} 

Our simulations were performed on the SCRI
cluster of fast RISC workstations. This research project was partially
supported by 
the Department of Energy under contract DE-FG05-87ER40319 and 
by the Supercomputer Computations
Research Institute which is partially funded by the U.S. Department of
Energy through Contract No. DE-FC05-85ER250000.  U. Hansmann is supported
by Deutsche Forschungsgemeinschaft under contract H180411-1.

\end{document}